\input epsf
\documentstyle[prl,floats,aps,psfig]{revtex}
\tighten
\begin{document}
\draft
\twocolumn[\hsize\textwidth\columnwidth\hsize\csname @twocolumnfalse\endcsname
\hfill\large{\bf preprint AEI-1999-25}%, gr-qc/9912017}
\bigskip\bigskip
%\preprint{AEI-1999-25, gr-qc/9912017}

\title{Pragmatic approach to gravitational radiation reaction in
binary black holes}
\author{Carlos O. Lousto
}
\address{Max-Planck-Institut f\"ur Gravitationsphysik,
Albert-Einstein-Institut, Am M\"uhlenberg 1, D-14476 Golm, Germany,\\
and 
Instituto de Astronom\'{\i}a y F\'{\i}sica del Espacio--CONICET,
Buenos Aires, Argentina
}
%\date{Received 8 October 1999; published 5 June 2000}
\maketitle

\begin{abstract}

We study the relativistic orbit of binary black holes in systems with
small mass ratio. The trajectory of the smaller object (another black
hole or a neutron star), represented as a particle, is determined by
the geodesic equation on the perturbed massive black hole spacetime.
The particle itself generates the gravitational perturbations leading
to a problem that needs regularization. Here we study perturbations
around a Schwarzschild black hole using Moncrief's gauge invariant
formalism. We decompose the perturbations into $\ell-$multipoles to show
that all $\ell-$metric coefficients are $C^0$ at the location
of the particle. Summing over $\ell$, to reconstruct the full metric, gives
a formally divergent result. We succeed in bringing
this sum to a generalized Riemann's $\zeta-$function regularization
scheme and show that this
is tantamount to subtract the $\ell\to\infty$ piece to each multipole.
We explicitly carry out this regularization and numerically
compute the first order geodesics. Application of this method to
general orbits around rotating black holes would generate
accurate templates for gravitational wave laser interferometric detectors.

\end{abstract}
\vskip2pc]

\section{Introduction}\label{sec:Introduction}

The 
computation of the gravitational radiation generated by binary black holes
is of great theoretical and observational interest. On one hand these
binary systems are among the best candidates to be detected by the new
generation of gravitational wave detectors. On the other hand the
theoretically very interesting two body problem in General Relativity
remains an unsolved task for full numerical approaches.
An appropriate astrophysical model to compute gravitational radiation
coming from the capture of stars by massive black holes at the center
of galaxies is provided by the perturbative approach applied to binary
systems where one black hole is much more massive than the other.
In the perturbative regime, linearized
Einstein equations can be brought to two simple wave equations for the
two polarizations of the gravitational field\cite{RW57,Z70}.
This problem has been recently revisited to include initial data into the
formulation in order to be able to start numerical integrations from finite
separations of the holes\cite{LP97a}. It also proved to be an excellent arena
where to test the accuracy of the Longitudinal-conformally flat ansatz
to solve the initial value problem for binary black holes\cite{LP97b,LP98}.
In order to answer questions like what is the
the displacement of the innermost stable circular orbit (ISCO) and the
rate of increase of the characteristic gravitational frequency due to
inspiralling orbits,
it is important to go beyond the leading approximation.
Technically, one has to compute
the geodesic trajectory of a particle in the perturbed black hole spacetime
generated by the particle itself. Taking this effect into account
will make otherwise particle's bounded orbits actually to inspiral
towards the bigger black hole. This consistent description
of the first order perturbative approach opens the door to second order
perturbative studies\cite{CL99} which allow to compute
gravitational radiation from binary black holes with much higher accuracy
and for systems with not so small mass ratio.
Such studies would teach us about new nonlinear
physical effects as well as to produce accurate templates to analyze the
forthcoming data from ground (and space) based laser interferometric detectors.

Recent work on gravitational radiation reaction \cite{MST97,QW97}
prompted a renewed theoretical interest on this problem.
The approach we will develop here privileges its direct computational
implementability.
Throughout this paper we will use results and techniques of
Ref.\ \cite{LP97b}. There it was computed the gravitational
radiation generated by the collision of two non-spinning black holes,
one much less massive than the other, starting from rest at a finite
distance. The extreme mass ratio allowed to describe the problem of
gravitational radiation as
perturbations about the Schwarzschild metric. Hence,
instead of working with all ten metric perturbations, seven even parity
and three odd parity (that identically vanish for our axially symmetric
problem), the relevant perturbative information
is organized into (one in the even parity case, two in general) the Moncrief
waveform\cite{M74}
\begin{eqnarray}
\psi_\ell (r,t)&=&\frac r{{\lambda }+1}
\left[ K^\ell+\frac{r-2M}{\lambda r+3M}\left\{
H_2^\ell-r\partial K^\ell/\partial r\right\} \right]
\label{psidef}
\end{eqnarray}
where we have used Zerilli's\cite{Z70} normalization for $\psi_\ell $ and
notation for ${\lambda }\doteq (\ell +2)(\ell -1)/2\ .$

One of the advantages of working with $\psi_\ell$ is its gauge
invariance under first order diffeomorfisms. This allows us to choose any
convenient gauge, like the Regge-Wheeler gauge\cite{RW57}, to make
computations. $\psi_\ell $ satisfies  a single (in
general two, with the odd parity case) wave equation
\begin{equation}
-\frac{\partial ^2\psi_\ell }{\partial t^2}+
\frac{\partial ^2\psi_\ell }{\partial r*^2}%
-V_\ell (r)\psi_\ell ={\cal S}_\ell (r,t)\ ,  \label{rtzerilli}
\end{equation}
where $r^{*}\equiv r+2M\ln (r/2M-1)$, $V_\ell $ is the Zerilli potential,
and ${\cal S}_\ell (r,t)$, is the source term generated by the small hole,
given in Ref.\ \cite{LP97a}.
The smaller hole is described as a point particle of proper mass $m_0$, its
stress energy tensor given by 
\begin{equation}
T^{\mu \nu }=m_0\frac{U^\mu U^\nu }{U^0r^2}\delta [r-r_p(t)]\delta ^2[\Omega
]\ ,  \label{tmunu}
\end{equation}
where $U^\mu $ is the particle 4-velocity. The two dimensional delta
function $\delta ^2[\Omega ]$ gives the angular location of the particle
$(\theta _p,\phi _p)$.
Since $T^{\mu \nu }$, and hence ${\cal S}_\ell (r,t)$ are already proportional
to $m_0$, to first perturbative order, the radial trajectory
$r_p(t)$, follows from the geodesic
equation in the background geometry (Schwarzschild's here).
There are situations when one
wants to know the trajectory of the small hole to the next order. In the
computation of the gravitational radiation to second perturbative order one
needs to know the source term to second order, hence the trajectory of the
hole on the first order metric (background plus first order perturbations).
Even within first order perturbation theory one would also like
to go further in order to compute secular effects like the
particle's bounded orbit decay around a much bigger black hole.

For the sake of simplicity we will treat here the
radial infall of a particle into a nonrotating hole. This problem contains
many of the relevant features that occur for more general orbits. 
To first perturbative order, the trajectory of the
particle is given by a geodesic on the first order metric (Schwarzschild
plus first order perturbations). This is so because the only ''forces''
acting on the particle are gravitational. The time component of the particle's
four-momentum, $P_t=m_0g_{tt}(dt/d\tau )\doteq -m_0E,$ is no longer a
conserved quantity along the
trajectory. We then have to deal with the radial and time (here the only
nontrivial ones) components of the geodesic equation.
We combine these two equations into a single equation of
motion for $r(t)$ by eliminating the dependence on the geodesic parameter.
Linearization of the resulting equation and subtraction of the zeroth order
geodesic gives the deviation of the trajectory $\Delta r_p(t)$ from the
zeroth order one
$
\dot r_p(t)=\partial_t z_p=-(1-2M/z_p)\sqrt{\frac{2M/z_p-2M/z_0}{1-2M/z_0}},
$
directly in terms of Schwarzschild coordinates, ready for further applications
\begin{eqnarray}
\Delta\ddot r_p=A\ \Delta r_p+B\ \Delta\dot r_p+C
\end{eqnarray}
where
\begin{eqnarray}\label{C}
A&=&\frac{2M}{r^3}\left[3-\frac{3M}{r}
-\frac{(1-3M/r)\dot r_p^2}{(1-2M/r)^{2}}\right],\nonumber\\ 
B&=&\frac{6M\dot r_p}{r^2(1-2M/r)},\nonumber\\
C&=&\left[
\stackrel{(1)}{\Gamma^t}_{rr}\dot r_p ^3+
\big(2\stackrel{(1)}{\Gamma^t} _{tr}-\stackrel{(1)}{\Gamma^r} _{rr}\big)
 \dot r_p ^2+\big(\stackrel{(1)}{\Gamma^t} _{tt}-
2\stackrel{(1)}{\Gamma^r} _{tr}\big) \dot r_p \right.\nonumber\\
&&\left.
-\stackrel{(1)}{\Gamma^r}_{tt}\right]=\sum_{l=0}^\infty C_\ell.
\end{eqnarray}

The (numerical) integration of this expression gives the trajectory of the
particle correct to first perturbative order, $\left( m_0/M\right) $.
Since the metric perturbations %(each of its $\ell-$ multipoles)
generated by a particle seem to involve radial delta functions
(and derivatives of them) evaluated at the location of the particle
(see Eqs.\ (\ref{Kextract})-(\ref{H1}) below),
the first problem to face
here is to evaluate the connection coefficients at $r_p$. We will next
show explicitly that the metric is actually $C^0$ at the location of
the particle. Connection coefficients have a finite
jump and they can be computed as the average of its values at
$r_p\pm\epsilon$ with $\epsilon\to0$. While this accounts for the 
terms coming from the radial 
Dirac delta in Eq.\ (\ref{tmunu}), as we will see, the angular delta needs 
to be explicitly regularized by a different method.

\section{Continuity of metric coefficients}

The $tt$ component of Einstein's equations give us the Hamiltonian
constraint. In the Regge-Wheeler gauge $(h_1=h_0=G=0)$
it is given by Zerilli's\cite{Z70} Eq.\ (C7a).
Only two metric coefficients $(K^\ell$
and $H_2^\ell)$ appear in this equation and none of its time derivatives. We
recall here that metric coefficients have an explicit multipole index $\ell$
since we have decomposed their angular dependence into tensor harmonics.
Consequently
$K^\ell$ and $H_2^\ell$ are only functions of $t$ and $r$ .
Taking the Regge-Wheeler gauge as an intermediate step, the definition of
$\psi_\ell $, and the Hamiltonian constraint
we can express these two metric coefficients in
terms of $\psi_\ell$ only (and source terms) 
\begin{eqnarray}
K^\ell&=&\frac{6M^2+3M\lambda r+\lambda (\lambda +1)r^2}
{r^2(\lambda r+3M)}\psi_\ell
+\left( 1-\frac{2M}r\right) \,\partial_r \psi_\ell\nonumber\\
&&-\frac{%
\kappa \ U^0(r-2M)^2}{(\lambda +1)(\lambda r+3M)r}\delta [r-r_p]\ .
\label{Kextract}
\end{eqnarray}
and
\begin{eqnarray} \label{H2extract}
&&H_2^\ell=-\frac{9M^3+9\lambda M^2r+
3\lambda ^2Mr^2+\lambda ^2(\lambda +1)r^3}{%
r^2(\lambda r+3M)^2}\,\psi_\ell   \nonumber\\
&&+\frac{3M^2-\lambda Mr+\lambda r^2}{r(\lambda
r+3M)}\partial _r\psi_\ell +(r-2M)\partial _r^2\psi_\ell \nonumber \\
&&+\frac{\kappa U^0(1-2M/r)[\lambda ^2r^2+2\lambda Mr-3Mr+3M^2]}{%
(\lambda +1)(\lambda r+3M)^2}\delta [r-r_p]\nonumber\\
&&-\frac{\kappa U^0(r-2M)^2}{
(\lambda +1)(\lambda r+3M)}\delta' [r-r_p]\ .  
\end{eqnarray}

Integration over $r$ of the Hamiltonian constraint tells us that 
the leading behavior is given by
\begin{equation}
\partial _r\psi_\ell \sim 
\kappa\ U^0(r-2M)\delta [r-r_p]/(\lambda +1)/(\lambda r+3M). 
  \label{psiprima}
\end{equation}
This can be used to prove that the metric coefficients at the location of the
particle are actually $C^0$, by taking up to second derivatives and using
the equation above to cancel derivatives of the Dirac's delta.
The same $C^0$ behavior at $r_p$ can be proven for $H_1^\ell$.
we now consider the $tr$ and $t\theta $ (or $t\varphi)$ components of the
Einstein equations that give us the momentum constraint. In the
Regge-Wheeler gauge they are given by Zerilli's\cite{Z70} Eqs.\ (C7b)
and (C7d). We combine them to eliminate the dependence on $H_1^\ell$ , and
obtain after integration over $r$ that
$%\begin{equation}
\partial _r(\partial_t{\psi_\ell })\sim -
\kappa \ U^0\stackrel{.}{r}_p(r-2M)\partial _r\delta [r-r_p]/
(\lambda +1)/(\lambda r+3M).
\label{psipuntoprima}
$%\end{equation}

From Zerilli's\cite{Z70} Eq. (C7b) and the expressions for
$\partial_t{K^\ell}$
and $\partial_t{H}_2$ in terms of $\partial_t{\psi_\ell }$, we find the last 
metric coefficient in the Regge-Wheeler gauge 
\begin{eqnarray}\label{H1}
&&H_1^\ell =r\partial _r(\partial_t{\psi_\ell })+\frac{\lambda r^2-3M\lambda
r-3M^2}{%
\left( r-2M\right) (\lambda r+3M)}\partial_t{\psi_\ell }- \\
&&\frac{\kappa \ U^0\stackrel{.}{r}_p(\lambda r+M)}{(\lambda +1)(\lambda
r+3M)}\delta [r-r_p]
+\frac{\kappa \ U^0\stackrel{.}{r}_pr(r-2M)}{(\lambda
+1)(\lambda r+3M)}\delta' [r-r_p]. \nonumber
\end{eqnarray}

This equation together with $H_0^\ell=H_2^\ell$ (valid for head-on collisions),
(\ref{H2extract}), and (\ref{Kextract}) give us all metric
perturbations on the chosen hypersurface in
terms only of $\psi_\ell $ and $\partial_t{\psi_\ell }$ (and the source).
We recall
here that since our case has axial symmetry only even waves are
generated.

\section{Regularization}

\begin{figure}
\epsfysize=3.0in \epsfbox{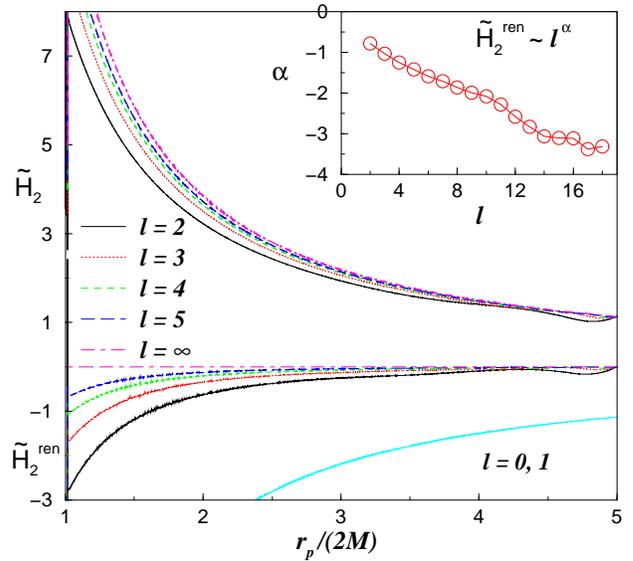}
\caption{The upper part of the plot shows the metric coefficient
$\tilde{H}_2^\ell$
along the trajectory of the particle in units of $m_0$.
Curves with
increasing $\ell$ quickly tend to superpose to that for $\ell\to\infty$.
The lower part of the plot represents the renormalized metric
$\tilde{H}_2^{\ell\ ren}=\tilde{H}_2^\ell-\tilde{H}_2^\infty$.
The inset figure gives the leading decay power, $\alpha$,
of ${H}_2^{\ell\ ren}$ at a fiducial $r_p$.
%, approaching $\approx-3$ for large $\ell$.
}\label{fig:H2}
\end{figure}

In Fig\ \ref{fig:H2} we plot the results of computing by means of
Eq.\ (\ref{H2extract}) the metric coefficient $\tilde{H}_2\dot=(1-2M/r_p)
\sqrt{2\ell+1}H_2^\ell$ along the
trajectory of the particle $r_p(t)$. It clearly shows that it is finite
at each point of the trajectory. The other notable feature is that
for large $\ell$ the curves quickly accumulate over the $\ell\to\infty$
asymptotic curve. Since one has then to sum over all $\ell$
contributions, this sum clearly diverges. One way of regularizing this
sum is to subtract to each mode precisely the $\ell\to\infty$ contribution,
and then verify the convergence of the remanent series.
The result of such regularization is shown in the lower half of
Fig\ \ref{fig:H2}. The same qualitative results are found for the other metric
coefficients $H_1^\ell(r_p),\ K^\ell(r_p),$ and
$H_0^\ell(r_p)=H_2^\ell(r_p)$. Non-radiative multipoles $\ell=0,1$
can be found analytically and represent the mass and linear
momentum contributions of the particle to the system \cite{Z70}.
The regularization just described is ambiguous up to a finite piece. To
give a definite prescription we note that it can be brought into the
generalized Riemann's $\zeta-$function\ \cite{BD82} procedure as follows.
The numerical
behavior of all metric coefficients shows that they can be decomposed
into two pieces: One that generates the finite behavior for $\ell\to\infty$
and the other strongly decaying for large $\ell$ (labeled with a
$ren$ below). Thus, for instance, for
the perturbed metric component $g_{rr}$ we can write
\begin{eqnarray}
&&H_2(t,r,\theta,\phi)=\sum_{\ell m}H_2^{\ell m}(t,r) Y_{\ell m}(\theta,\phi)\\
&=&\sum_{\ell=0}^\infty\left\{(2\ell+1-\beta)^{-\beta} H_2^\infty +
H_2^{\ell\ ren}\right\}\sqrt{\frac{2\ell+1}{4\pi}}P_\ell(\theta)\nonumber
\end{eqnarray}
where in the second equality
we made use of the axial symmetry of the problem and we have chosen
the $\beta-$parametrization motivated by the $D-$dimensional
extension of the conformally flat initial value problem, where $\beta=4-D$.
When we evaluate this metric coefficient at the location of the particle
we find
\begin{eqnarray}
&&H_2(t,r_p(t),\theta_p=0)=\\
&&2^{-\beta+1/2}\frac{H_2^\infty}{\sqrt{4\pi}}\zeta(\beta-1/2,1/2)
+\sum_{\ell=0}^\infty \sqrt{\frac{2\ell+1}{4\pi}}H_2^{\ell\ ren}.\nonumber
\end{eqnarray}
where the Riemann's $\zeta$-function is $\zeta(a,b)=\sum_{\ell=0}^\infty
(\ell+b)^{-a}$.
Numerically, we observe that $\beta=1/2$ in order to lead to the
finite $\ell\to\infty$ behavior. Since the analytically continued
$\zeta$-function gives $\zeta(0,1/2)=0$ we must subtract to each
multipole just the $\ell\to\infty$ piece.
The renormalized metric coefficients that enter into the head-on geodesic
equation $(H_0^{\ell\ ren}=H_2^{\ell\ ren}, H_1^{\ell\ ren})$ vanish for
large $\ell$ as $\sim\ell^{-3}$, as we numerically roughly estimated
(cf. Fig.\ \ref{fig:H2}).
This implies that the regularized connection coefficients scale
as $\sim\ell^{-2}$ for large $\ell$ and that its sum, to compute $C$,
up to a finite maximum $L$ (as is done in practice) scale as $\sim L^{-1}$.
This can be used to estimate the error produced by a truncation of the
series. 

\begin{figure}
\begin{center}
\begin{tabular}{@{}lr@{}}
\psfig{file=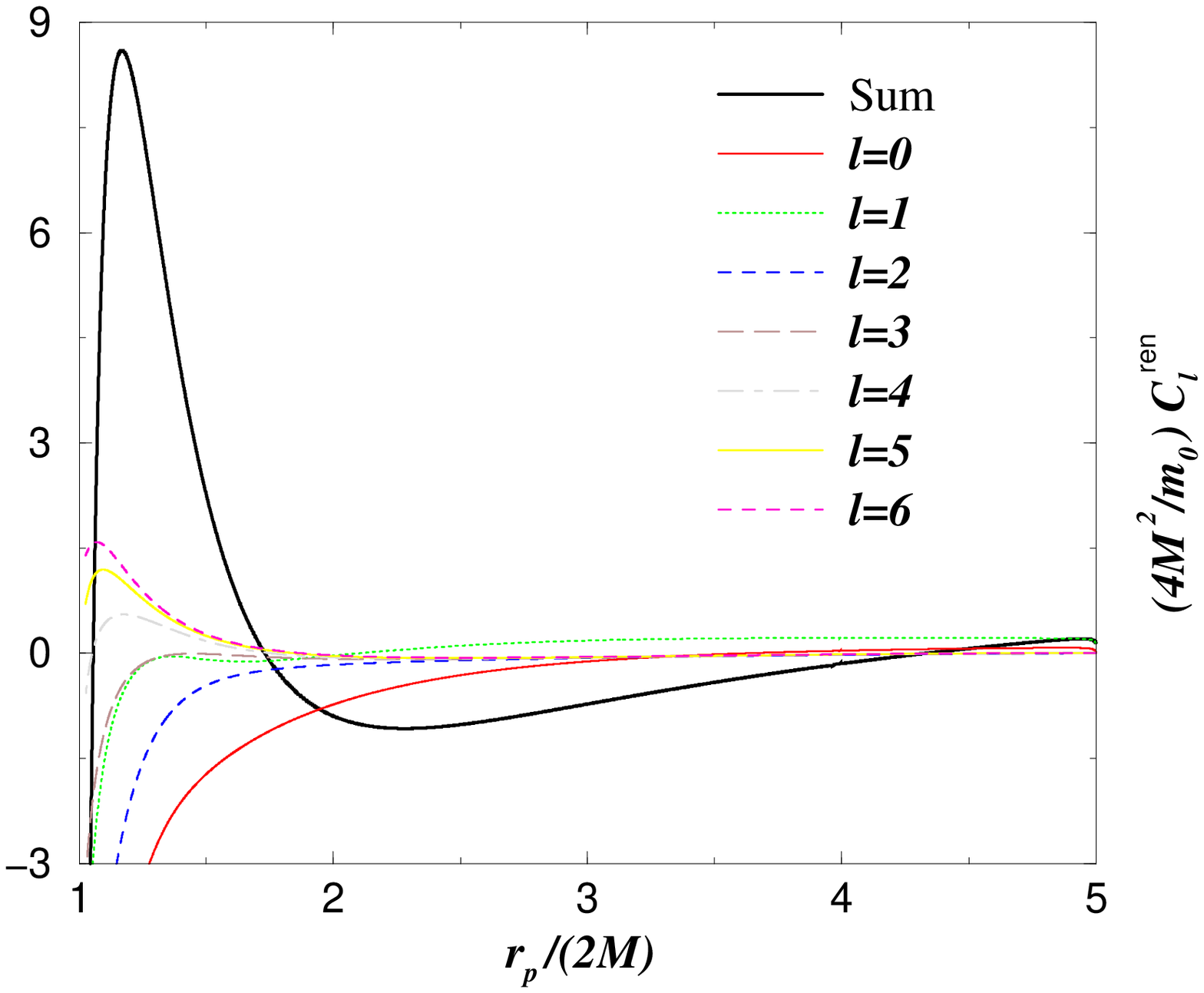,width=90mm,clip=}\\
\psfig{file=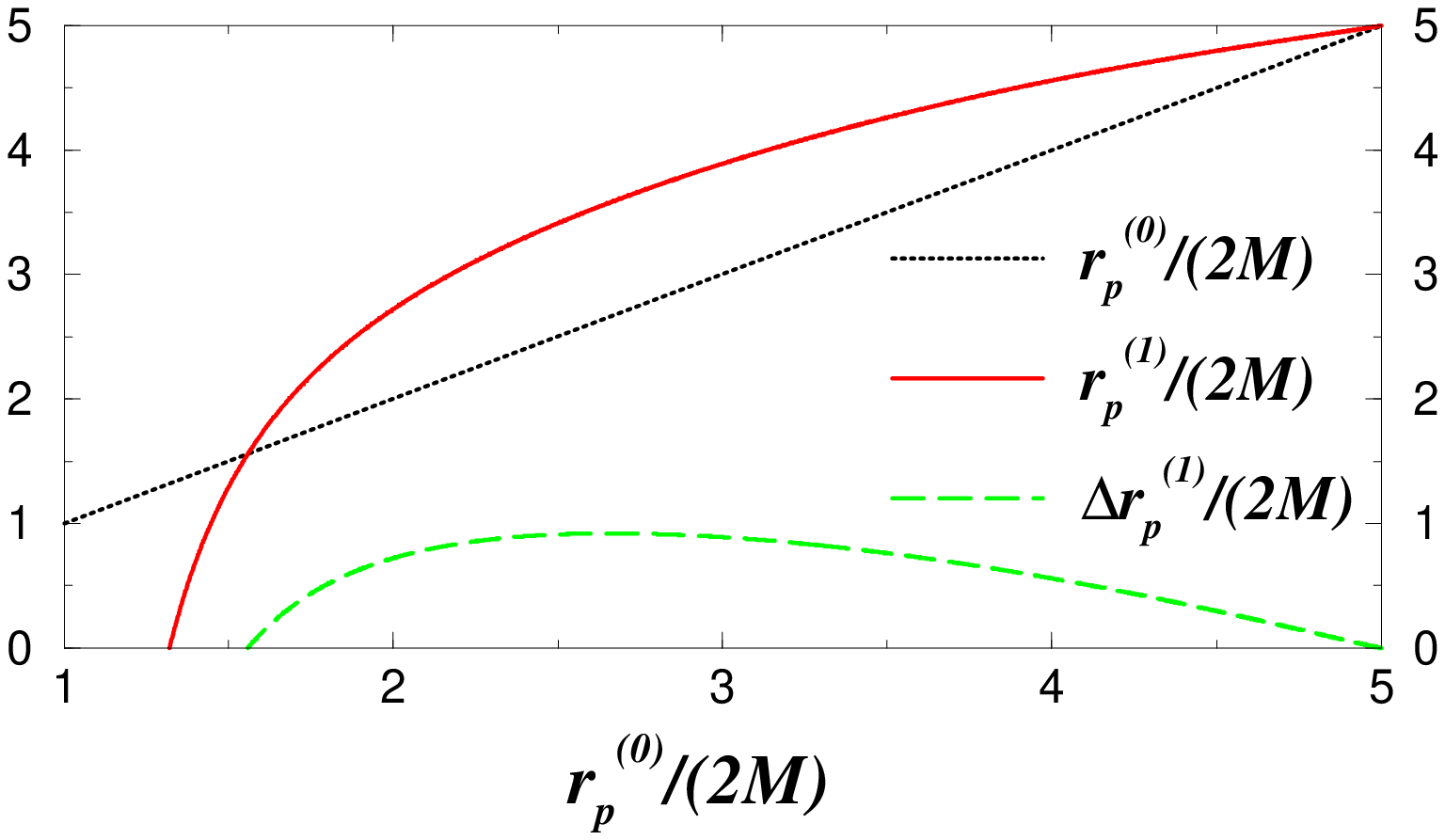,width=85mm,clip=}\\
\end{tabular}
\end{center}
\caption[]
{The renormalized radiative piece of the reaction on $\ddot r_p$.
We show the regularized sum over $\ell\leq10$ multipole contributions
to $C$, as defined in Eq.\ (\protect\ref{C}).
Below the first order trajectory $r_p^{(1)}$, for $m_0=0.1M$,
is compared to the zeroth order one, $r_p^{(0)}$.
}
\label{fig:Cren}
\end{figure}

From the metric we build up the connection coefficients that enter into
the geodesic equation. The piece of the relative acceleration that is
summed over $\ell$, denoted by $C$ in Eq.\ (\ref{C}), needs to be
regularized in the same way metric coefficients do. This piece summed
over the $\ell\leq10$ radiative multipoles is plotted in Fig.\ \ref{fig:Cren}
$(\ell=0$ and $\ell=1$
multipoles have been solved analytically using the additional
gauge freedom to set $H_2=K$ and, in the $\ell=0$ case, also $H_1=0$).
We find the radiation reaction effects can be qualitatively described
in two regimes. First, the renormalized acceleration is very small in
the early stages of the trajectory. This is expected for a
particle starting at rest, since the gravitational radiation
is dominated by the bremsstrahlung mechanism.
%This regime can be directly
%compared with post-Newtonian predictions using the quadrupolar formula.
Then, radiation reaction effects become more important as the particle
approaches the maximum of the Zerilli's potential (around
$r_{max}\approx3.1M)$. They tend to decelerate the particle
% closer to the bigger black hole
with respect to the zeroth order
(Schwarzschild) geodesics. These is what one would qualitatively
expect a priori since the system is
loosing energy and momentum in the form of gravitational radiation.
Another important feature is that the radiation reaction effect is
low $\ell-$dominated (only after renormalization), but still the sum over
large $\ell$ give an important contribution, and one has to consider
higher multipoles than one usually takes into account when computing,
for instance, the total energy radiated reaching infinity.

A related approach to the one presented here has been developed
independently\cite{BO00} and applied to scalar radiation.
It considers the large $\ell$ dependence of
the tail term of the reaction force\cite{QW97}, and regularizes the
sum over $\ell$ modes by
subtracting the non-convergent terms plus a finite part to be determined.
Applied to the equivalent case we treated here the finite part
to be subtracted vanishes and both procedures coincide.
Besides, this method was implemented to study static scalar and Electric
fields in the Schwarzschild background \cite{B00}, 
leading to the correct known expressions of the self force.
It would be very interesting to
compare the results of our procedure with still other methods of
regularization in order to cross check for possible ambiguities in
the determination of the remaining finite parts.

\section{Discussion}

Here we reported on a first important step to improve our ability to compute
gravitational radiation from binary black holes. The next step is
to apply the $\zeta-$function regularization method to general bounded
orbits in the Schwarzschild background and compare its results with the
energy and momentum balance estimates.
Besides, our renormalization procedure is explicitly
gauge invariant since we use Moncrief's waveform.
 Another key problem to attack is the
orbiting particle in a rotating black hole background. %\cite{H99}.
Perturbations of
the Kerr metric are described by the Teukolsky equation and can be decomposed
into multipoles (via spin-weighted spheroidal harmonics) in the frequency
domain, i.e. after a Fourier decomposition of the time dependence. This
property would still allow to apply our regularization scheme and
compute the decay of bounded orbits consistently to first order.
Besides, it would open the possibility to study
second order perturbations\cite{CL99} and
obtain a remarkable improvement in our ability to compute gravitational
radiation from binary systems with not so small mass ratio. This will
not only be of use for LISA's detection of gravitational radiation from
black holes in the center of nearby galaxies, but also relevant for
ground based interferometers, sensible to frequencies corresponding to
black hole / neutron star binaries.

\begin{acknowledgments}
%{\it Acknowledgments}:
It is my pleasure to acknowledge very helpful discussions on this work
with L. Burko, C. Cutler, A. Ori, R. Price, and B. Schutz.
C.O.L. is a member of the Carrera del Investigador Cient\'\i fico 
of CONICET, Argentina.
\end{acknowledgments}

\end{document}